\documentclass[11pt]{article}
\usepackage{amsmath,epsf,amssymb,latexsym,enumerate,cite,shadow,array,color}
\usepackage[ps,dvips,matrix,arrow,frame,import,curve,color]{xy}
\newif\iffigs\figstrue

\DeclareFontFamily{U}{rsf}{}
\DeclareFontShape{U}{rsf}{m}{n}{
  <5> <6> rsfs5 <7> <8> <9> rsfs7 <10-> rsfs10}{}
\DeclareMathAlphabet\Scr{U}{rsf}{m}{n}

\def\pplogo{\vbox{\kern-\headheight\kern -29pt
\halign{##&##\hfil\cr&{
\ppnumber}\cr\rule{0pt}{2.5ex}&\ppdate\cr}
}}
\makeatletter
\def\ps@firstpage{\ps@empty \def\@oddhead{\hss\pplogo}%
  \let\@evenhead\@oddhead 
}
\def\maketitle{\par
 \begingroup
 \def\thefootnote{\fnsymbol{footnote}}
 \def\@makefnmark{\hbox{$^{\@thefnmark}$\hss}}
 \if@twocolumn
 \twocolumn[\@maketitle]
 \else \newpage
 \global\@topnum\z@ \@maketitle \fi\thispagestyle{firstpage}\@thanks
 \endgroup
 \setcounter{footnote}{0}
 \let\maketitle\relax
 \let\@maketitle\relax
 \gdef\@thanks{}\gdef\@author{}\gdef\@title{}\let\thanks\relax}
\makeatother

\newcommand{\bea}{\begin{eqnarray}}
\newcommand{\eea}{\end{eqnarray}}
\newcommand{\be}{\begin{equation}}
\newcommand{\ee}{\end{equation}}

\newcommand{\rf}[1]{(\ref{#1})}

\begin{document}

\thispagestyle{empty}

\begin{titlepage}

\hskip 1cm 

\vskip  2 cm

\vspace{24pt}

\begin{center}
{ \LARGE \bf{     An Update on Perturbative  N=8 Supergravity
 }}

\vspace{40pt}

{\bf Renata Kallosh}

\

 \textsl{Department of Physics,
    Stanford University}\\ \textsl{Stanford, CA 94305-4060, USA}

\vspace{2cm}

\end{center}

\begin{abstract}

 According to the recent pure spinor analysis  of the UV divergences by Karlsson,  there are no divergent 1PI structures beyond 6 loops in D=4  N=8 supergravity.  In combination with the common expectation that the UV divergences do not appear at less than 7 loops, this may imply that the 4-point amplitude in D=4  N=8 supergravity is all-loop finite.  This differs from the result of the previous studies of pure spinors, which suggested that there is a UV divergence at 7 loops in D=4.  Therefore an independent investigation of the pure spinor formalism predictions is desirable, as well as continuation of explicit loop computations. In the meantime, we revisit here our earlier arguments on UV finiteness of N=8 supergravity based on the absence of the off-shell light-cone superspace counterterms, as well as on the $E_{7(7)}$ current conservation. We believe that both arguments remain valid in view of the developments in this area during the last few years.

\end{abstract}

\end{titlepage}

\newpage

\section{Introduction}

N=8 D=4 supergravity \cite{Cremmer:1979up} remains a puzzle: the perturbative UV properties of this theory  were revisited many times over the years. From the very beginning there was a hope that it is perturbatively UV finite. However, already in 1980 the on-shell supersymmetric counterterms were constructed  \cite{Kallosh:1980fi,Howe:1980th}, using the N=8 on-shell superspace \cite{Brink:1979nt}. The existence of the linearized 3-loop on-shell supersymmetric invariant 
\cite{Kallosh:1980fi,Howe:1981xy} was interpreted as an expectation that the 3-loop UV divergence is possible. Therefore it came as a shock to supergravity experts  that in actual computations in  \cite{Bern:2007hh} the cancellation of all UV divergent 3-loop terms was discovered. This cancelation was later explained by two different methods: the first was using the light-cone analysis in \cite{Kallosh:2009db}, and later, using 
$E_{7(7)}$ symmetry in \cite{Brodel:2009hu},  \cite{Elvang:2010jv}.
In all cases an agreement was reached that the first UV divergence is possible at 7 loops,  the earliest. 
 
The explicit maximal supergravity calculations suggest that  the absence of UV divergences is consistent with the maximal supersymmetric Yang-Mills type formula 
\be
D < 4 + {6\over L} \, , \qquad   L>1
\label{zvi}\ee 
for the first 4 loops, as  proposed in \cite{Bern:1998ug}. The formula means that at the loop order $L$ the theory in $D$ dimensions $D < 4 + {6\over L}$ is finite and it starts having divergences at the critical dimension $D_c = 4 + {6\over L}$. 
If this formula is valid at higher loops, maximal supergravity would be UV finite to all orders.
However, it was suggested in \cite{Bjornsson:2010wm} that 
at 5 loops there is a radical change. A new formula determining the critical $D$ for a given $L$ for N=8 supergravity was derived using the pure spinor formalism:
\be
D^{\rm BG}< 2 + {14\over L}\, , \qquad L > 4 \ .
\label{BG}\ee
The critical dimension where the theory starts to have UV divergences is $D_c= 2 + {14\over L} $.
This means that the 5-loop critical dimension is $D_c = 24/5$ instead of the Yang-Mills type one, $D_c = 26/5$. If the 5-loop explicit computations were to confirm $D_c = 24/5$, it would also mean, according to \rf{BG}, that N=8 supergravity is UV divergent at 7 loops in D=4 since
\be
D_c^{\rm BG}(L=7) = 2 + {14\over 7}=4 \ .
\ee
The bound \rf{BG} was also derived in a version of pure spinor formalism in D=11 supergravity in \cite{Cederwall:2012es}, and confirmed by an application of the maximal unitarity cut method in \cite{Bern:2014sna}. It was, however, noticed in   \cite{Bern:2014sna} that the maximal cut method by itself, valid for individual diagrams in a double-copy computations, does not explain why the complete set of diagrams has no UV divergence in many cases.

During the last few years there was a hope that an explicit 5-loop computation in N=8 supergravity will be able  to distinguish between these two cases and put things to rest: if $D_c$ is found to be  24/5 it would mean L=7 UV divergence and the finiteness conjecture will be falsified. If $D_c$ is found to be 26/5, more work will be required to study higher loops, $L>7$.
 However, the computation is hard and has not yet been accomplished: it would be great to see the answer! Meanwhile,  the new study of the pure spinor formalism in \cite{Karlsson:2014xva} makes the 5-loop paradigm questionable.
 
\section{ Pure spinor superspace}
In both studies of pure spinors in application to N=8 supergravity in \cite{Bjornsson:2010wm} and \cite{Cederwall:2012es} a certain assumption was made, which was challenged by Karlsson in the recent paper \cite{Karlsson:2014xva}. The pure spinor superspace in D=11 supergravity in addition to $ \theta^\alpha$ spinors has some bosonic spinors $\lambda^\alpha$  which are pure:  they satisfy the constraint $\lambda\gamma_a \lambda =0$, where $a=0,...,10$. An additional  set of constrained  spinors $(\bar \lambda_\alpha, r_\alpha)$, which are  counterparts to $( \theta^\alpha, \lambda^\alpha)$,
 satisfying the constraints $\bar \lambda \gamma_a \bar \lambda =0$, $ \bar \lambda \gamma_a r =0$,  is also required \cite{Berkovits:2005bt},\cite{Cederwall:2010tn}. In analogous case of maximal  supersymmetric YM theory the number of components in $r$ is 11, whereas in D=11 supergravity, the number of  components in $r$ is 23.
 
 According to  \cite{Karlsson:2014xva}, the assumption in \cite{Bjornsson:2010wm} and \cite{Cederwall:2012es} that in each case the relevant UV divergences have a maximal dependence on $r$, namely 11 for YM and 23 for supergravity, has to be revisited. She has found that in YM case the dependence does not exceed 10 and in supergravity 15. Therefore in the YM case the difference is small,  the critical dimension formula \rf{zvi} remains intact. However, in supergravity the difference leads to dramatic consequences. 
 
The change to equation \rf{BG}, which is now $D> 2 + {8\over L} \, ,  L\geq 2$, by itself is not very illuminating: it shows that the UV divergence may be present in 5-loops D=4, N=8.  It means, as pointed out in \cite{Karlsson:2014xva}, that the critical dimension formula in this case is not  useful. A different aspect of the analysis  makes an important point: she has established that 1PI contributions to UV divergences of the 4-point amplitudes are not available at $L > 6$  (and for the 5-point amplitude beyond $L > 7$). This means that no new UV divergences are possible in the 4-point amplitude besides the ones up to 6 loops.

But now we can apply the argument in \cite{Elvang:2010jv},  \cite{Kallosh:2009db} that the earliest UV divergences  for D=4, N=8 supergravity are at 7 loops. We end up, repeating after  \cite{Karlsson:2014xva}, that 4-point D=4, N=8 supergravity is UV finite.
 
If true, the result is spectacular. There is a large community of experts working in pure spinor formalism, see e.g. \cite{Berkovits:2005bt},
\cite{Bjornsson:2010wm}, \cite{Berkovits:2014rpa},
and a recent review \cite{Cederwall:2013vba}. If indeed the point of $r$-dependence spotted in \cite{Karlsson:2014xva} is correct, it can be tested. This will move the field in one or another direction. 

 In the meantime, before the consensus is reached with respect to the pure spinor formalism predictions and/or explicit results of the 5-loop computations are available, we would like to revisit the present status of other approaches to these issues.

\section{Revisiting the light-cone and $E_{7(7)}$ finiteness arguments}

\subsection{Light-cone superspace}

 Soon after the discoveries in \cite{Bern:2007hh}, 
we have combined in \cite{Kallosh:2009db} the helicity formalism used in supergravity amplitude constructions \cite{Bianchi:2008pu}
 with the analysis of the  counterterms in the light-cone superspace \cite{Brink:1982pd}.   This led to a realization that there are no local  off-shell light-cone superspace counterterms. As a result, we have argued in \cite{Kallosh:2010kk} that N=8 supergravity is perturbatively UV finite. 

So far, no direct omission was found in the arguments presented in \cite{Kallosh:2009db,Kallosh:2010kk}. This is not entirely surprising:  because of the absence of experts working in both fields simultaneously, in the  light-cone superspace formalism in supergravity and in the 
helicity formalism for amplitudes,  it was difficult to expect that the argument in \cite{Kallosh:2010kk} will be tested soon. The only indirect argument against it was that the conclusions of \cite{Kallosh:2009db,Kallosh:2010kk} did not match the tentative conclusions based on an assumption  that the off-shell harmonic superspace in N=8 supergravity was available  \cite{Bossard:2009sy}. 

However, harmonic superspace is well defined \cite{Galperin:2001uw} only in the super-Yang-Mills theory, not in supergravity.   Still it was conjectured to exist, which served as the basis for many predictions regarding loop computations in extended supergravity, for example in \cite{Bossard:2011tq}. It turned out more recently that the harmonic superspace predictions  are in contradiction with  number of recent computations in $N< 8$ supergravity. For example, based on the harmonic superspace conjecture,  a prediction of UV divergence was made for D=4, L=3, N=4,   for D=4, L=4, N=5, and for the half-maximal supersymmetry in D=5, L=2 in \cite{Bossard:2011tq}.
 The actual computations during the last two years \cite{Bern:2014lha}, \cite{Bern:2014sna} have shown  the absence  of these UV divergences, in contradiction  with \cite{Bossard:2011tq}.  These cases  were  given a name  ``enhanced cancellation'' in \cite{Bern:2014sna}.  It is interesting that in the double-copy formalism in which the explicit computations are performed in \cite{Bern:2014lha}, \cite{Bern:2014sna} the individual diagrams are UV divergent in agreement with the maximal cut unitarity method, however, the total contribution from all diagrams cancels in a rather mysterious way.

To summarize this part: now, at the end of 2014, we would like to re-iterate our light-cone argument  about the absence of local counterterms. The only indirect challenge to this argument  made in \cite{Bossard:2009sy}  was never justified, and over time the credibility of the arguments based on harmonic superspace in supergravity went down. In any case, even  independently of the particular examples,  {\it the off-shell light-cone counterterms were never constructed despite some attempts to do so}.\footnote{L. Brink, private communication} The first observation that they do not exist was made in 2008 \cite{Kallosh:2010kk}, and they are still not available. The absence of the light-cone superspace counterterms for N=8 supergravity is  one way of explaining why the UV divergences present in the  individual diagrams cancel for the complete set.

We have also explained why the absence  of the off-shell light-cone counterterms is not in contradiction with the existence of the old on-shell Lorentz covariant counterterms constructed in  \cite{Kallosh:1980fi,Howe:1980th,Howe:1981xy}: being on-shell,  they are not eligible to describe UV divergences consistently  \cite{Chemissany:2012pf}.

\subsection{$E_{7(7)}$ current conservation}
 
 We also studied  the situation in a Lorentz covariant formulation where one could have expected  that the $E_{7(7)}$-symmetry plays an important role.  We described our findings in \cite{Kallosh:2011dp}.
Here is a short summary of the situation.  We have found that the duality symmetries like $E_{7(7)}$ require a subtle analysis compared with the `brute force' construction of the $E_{7(7)}$ on-shell superinvariants which we have proposed back in 1980. The subtlety boils down to the fact that  this kind of symmetry is deformed when the higher derivative local superinvariants are added to the 2-derivative classical action of N=8 supergravity. We made an argument that such a consistent deformation is not possible, which predicts that that the theory is perturbatively UV finite.

The counter-argument was suggested in \cite{Bossard:2011ij}, where it was proposed that such a consistent deformation, which deforms both $E_{7(7)}$ as well as the classical action, might  exist. To counter  this point we have made an observation \cite{Kallosh:2012yy} based on earlier studies of non-linear dualities in \cite{Carrasco:2011jv}. Namely, we have explained 
that  a claim in \cite{Bossard:2011ij}  is equivalent to an unproven conjecture that a new N=8 supergravity of the Born-Infeld type must exist. By studying the octonionic nature of  $E_{7(7)}$ we explained why the validity of this conjecture is very unlikely. 
We proposed that, maybe, the authors of \cite{Bossard:2011ij} will be able to substantiate their claim by providing such a new N=8  Born-Infeld supergravity. In such case the whole story would go into a new direction since the new theory, if available, would already have all higher derivative terms from the beginning.  Would such theory by itself be finite by construction, if it is unique? Everyone seems to agree  that constructing N=8 Born-Infeld supergravity is a tall order. Therefore at present {\it the criticism in  \cite{Bossard:2011ij}, based on the conjecture of existence of  N=8 Born-Infeld supergravity, is not  justified}.

\

During the last two years  the new explicit computations in case of N=4 and N=5 supergravities were performed \cite{Bern:2014lha}, \cite{Bern:2014sna}. The 3-loop N=4 UV finite case had one of these enhanced cancellations, the 4-loop has a UV divergence \cite{Bern:2013uka}, however, it was recognized as related to $U(1)$ anomaly represented by  3 different structures derived earlier in \cite{Carrasco:2013ypa}. Finally the 4-loop N=5 supergravity, which was expected  to have no anomalies, has recently shown another case of enhanced cancellations, and no UV divergence \cite{Bern:2014sna}. It is a rather impressive computation which shows that all 3 anomaly-related UV divergences of N=4 theory are precisely cancelled when the contribution due to additional states of N=5 supergravity is taken into account.

In conclusion, we believe that our arguments related to UV properties of N = 8 supergravity made in \cite{Kallosh:2009db,Kallosh:2010kk} and \cite{Kallosh:2011dp} still hold. We hope to learn more about the puzzling UV properties of extended D=4 supergravities. From all near future developments, the studies of the pure spinor formalism predictions look  promising:   the result will either confirm the current 24/5 to 26/5 paradigm, to be resolved by the future explicit 5-loop computations, or support the finiteness arguments of the 4-point amplitude. Still, in view of the complicated nature of the pure spinor formalism, light-cone superspace and $E_{7(7)}$-symmetry of N=8 supergravity it would be important  to see the results of the explicit 5-loop and higher loop computations.

\section*{Acknowledgments}

 We are grateful to Martin Cederwall for a discussion of the present status of the pure spinor superspace and to Zvi Bern and John Joseph Carrasco   for the discussion of current status of the explicit computations, including the nature of the enhanced cancellations.  This work is supported by the SITP and by the NSF Grant PHY-1316699 and by the Templeton foundation grant `Quantum Gravity Frontiers'.


\end{document}